\documentstyle[aps,prl,multicol,fancyheadings,rotate,epsfig,amsbsy,amssymb,amstex]{revtex}

\def\bbbc{{\mathchoice {\setbox0=\hbox{$\displaystyle\rm C$}\hbox{\hbox
to0pt{\kern0.4\wd0\vrule height0.9\ht0\hss}\box0}}
{\setbox0=\hbox{$\textstyle\rm C$}\hbox{\hbox
to0pt{\kern0.4\wd0\vrule height0.9\ht0\hss}\box0}}
{\setbox0=\hbox{$\scriptstyle\rm C$}\hbox{\hbox
to0pt{\kern0.4\wd0\vrule height0.9\ht0\hss}\box0}}
{\setbox0=\hbox{$\scriptscriptstyle\rm C$}\hbox{\hbox
to0pt{\kern0.4\wd0\vrule height0.9\ht0\hss}\box0}}}}
\begin{document}
\title{Quantum Phase Diagram of the $t$-$J_z$ Chain Model}
\author{C.D. Batista and G. Ortiz}
\address{Theoretical Division, 
Los Alamos National Laboratory, Los Alamos, NM 87545}
\date{Received \today }
\maketitle

\begin{abstract}
We present the quantum phase diagram of the one-dimensional $t$-$J_z$
model for arbitrary spin (integer or half-integer) and sign of the
spin-spin interaction $J_z$, using an {\it exact} mapping to a spinless
fermion model that can be solved {\it exactly} using the Bethe ansatz.
We discuss its superconducting phase as a function of hole doping
$\nu$. Motivated by the new paradigm of high temperature
superconductivity, the stripe phase, we also consider the effect the
antiferromagnetic background has on the $t$-$J_z$ chain intended to
mimic the stripe segments.
\end{abstract}


\vspace*{-0.3cm} 

\begin{multicols}{2}

\columnseprule 0pt

\narrowtext
\vspace*{-0.5cm}

{\it Introduction.} 
Phase diagrams of generic models of strongly interacting quantum 
particles are considered fundamental to understanding the complex
physical behavior of cuprate superconductors, heavy fermion, and 
related compounds. It is rare to encounter situations where
unambiguously these diagrams can be completely determined and only a
few exceptional cases are exactly solvable. It is a purpose of this
paper to show that the $t$-$J_z$ chain belongs to this latter class of
models.  

A new paradigm in superconductivity springs up as a consequence of the
growing body of experimental evidence suggesting that the quantum state
of high temperature superconductors is a striped phase. Unlike
conventional conductors where the charge carriers distribute in an
spatially homogeneous way, the stripe paradigm assumes that carriers
cluster into quasi one-dimensional (1$d$) channels. These channels act as
domain walls separating different antiferromagnetic (AF) domains. It is
remarkable that experiments are consistent with a spin ordering that is
$\pi$-shifted across the wall \cite{tranqua}, indicating the
topological character of these extended defects \cite{topo}.

Motivated by this new paradigm Ref. \cite{ours} argued that planar
models, with appropriate inhomogeneous magnetic terms, breaking
translational and local spin $SU(2)$ symmetries are appropriate to
understand neutron scattering and angle-resolved photoemission
spectroscopy experiments in cuprates \cite{experim}. It is interesting
to understand why spin anisotropies are relevant to obtain substantial
pair hole binding and whether the stripes themselves have important
superconducting fluctuations. The simplest representation of a stripe
segment is realized by a $t$-$J_{z}$ chain model.

In this Letter we study the quantum phase diagram of the $t$-$J_z$
chain for arbitrary spin and sign of $J_z$ by using an exact mapping to
an attractive spinless fermion model, and solve this problem  by using
the Bethe ansatz integral equations. We then consider the effect of the
AF boundaries on the stripe as an effective confining potential and
determine the resulting phase diagram. While a superconducting phase
exists in both cases, the superconducting region is more prominent in
the latter.  

{\it Model Hamiltonian.} The Hamiltonian representing the 1$d$
$t$-$J_{z}$ model with $L$ sites (equal to the length of the chain,
i.e., lattice constant $a$=1) and $M$ holes with open boundary
conditions (BC) \cite{Note0} (the thermodynamic limit, $L,M\rightarrow
\infty $ with $\nu=M/L$ finite, is performed at the end of the
calculation), for arbitrary half-integer spin $S$, is
$\hat{H}=\hat{T}+\hat{H}_{J_{z}}$ with
\begin{eqnarray}
\hat{T} &=&-t\!\!\!\sum\Sb \alpha =1 \\ \sigma \in [-S,S] \endSb ^{L-1}
\!\! \hat{T}_{\alpha,\sigma} \; , \; \hat{T}_{\alpha ,\sigma} = c_{\alpha
\sigma }^{\dagger }c_{\alpha +1\sigma }^{\;}+{\rm H.c.} \nonumber \ ,
\\ \hat{H}_{J_{z}} &=&J_{z}\sum_{\alpha =1}^{L-1}S_{\alpha
}^{z}S_{\alpha +1}^{z} \ , \;\; S_{\alpha}^{z} =
\!\!\!\!\!\sum\limits_{\sigma \in [-S,S]} \! \!\sigma \ c_{\alpha
\sigma }^{\dagger }c_{\alpha \sigma }^{\;} \ . \nonumber
\end{eqnarray}
Here, $c_{\alpha \sigma }^{\dagger}$($c_{\alpha \sigma }^{\;}$)
creates(annihilates) a fermion of $z$ spin-component $\sigma$
in a Wannier orbital centered at $\alpha$. The Hilbert space of
states corresponds to a constrained space with no doubly occupied sites
\cite{Note1}.

Consider the set of parent states, labeled by the string configuration
$\vec{\boldsymbol{\sigma}}$, with $M$ holes and $L-M$ quantum particles,
$|\Phi _{0}(\vec{\boldsymbol{\sigma}}) \rangle$, defined as 
\begin{equation} 
|\Phi_0(\vec{\boldsymbol{\sigma}}) \rangle = |
\underset{L-M}{\underbrace{\sigma_{1} \sigma_{2} \sigma_{3}  \cdots
\sigma_{L-M}}} \  \underset{M}{\underbrace{\circ _{\;}\!\!\circ \circ
\circ \circ \cdots}} \ \rangle \ ,
\end{equation}
where $\sigma_{\alpha}$ indicates the $z$-component of the spin of the
particle at site $\alpha$. These states are eigenstates of
$\hat{H}_{J_{z}}$ with energy
$E_{M}(\vec{\boldsymbol{\sigma}})=J_{z}\sum_{\alpha=1}^{L-M-1} \sigma
_{\alpha }\sigma _{\alpha +1}$, and $z$-component of the total spin
$S_{z}=\sum_{\alpha =1}^{L-M}\sigma _{\alpha}$. 

>From a given parent state one can generate a subspace of the Hilbert
space ${\cal M}(\vec{\boldsymbol{\sigma}})$ by applying the hopping
operators $\hat{T}_{\alpha ,\sigma }$ to the parent state and its
descendants, $|\Phi_{i}(\vec{\boldsymbol{\sigma}})\rangle$,
\begin{equation}
|\Phi_{1}(\vec{\boldsymbol{\sigma}})\rangle =\hat{T}_{L-M,\sigma} \
|\Phi_{0}(\vec{\boldsymbol{\sigma}})\rangle 
\end{equation}
or, in general, 
\begin{equation}
|\Phi_{i}(\vec{\boldsymbol{\sigma}})\rangle =\hat{T}_{\alpha ,\sigma }\
|\Phi_{j}(\vec{\boldsymbol{\sigma}})\rangle \ .
\end{equation}
The dimension ${\cal D}$ of the subspace ${\cal
M}(\vec{\boldsymbol{\sigma}})$ is ${\binom{L}{M}}$. Moreover, these
different subspaces are orthogonal and are not mixed by the Hamiltonian
$\hat{H}$, although they can be degenerate. In the following we will
only consider the AF case $J_z > 0$. At the end we will discuss two
important generalizations: the ferromagnetic (FM) $J_z < 0$ and the
arbitrary integer spin hard-core boson cases. 

Among all possible initial configurations the one corresponding to the
N\'eel string $\vec{\boldsymbol{\sigma}}_0$ (i.e., $\sigma_{\alpha} =
(-1)^{\alpha} \ S$), which is two-fold degenerate, turns out to be
special.  We want to show now that for a given number of holes $M$ the
subspace generated by the N\'{e}el parent state, ${\cal
M}(\vec{\boldsymbol{\sigma}}_0) \equiv {\cal M}_{0}$, contains the
ground state. To this end, one has to realize that the kinetic energy
matrix elements $\langle \Phi_{i} (\vec{\boldsymbol{\sigma}})|\hat{T}|
\Phi_{j} (\vec{\boldsymbol{\sigma}})\rangle$ are the same for the
different subspaces ${\cal M}$. Nonetheless, the magnetic matrix
elements $ \langle \Phi _{i}(\vec{\boldsymbol{\sigma}})|\hat{H}_{J_{z}}
| \Phi_{j}(\vec{\boldsymbol{\sigma}})\rangle =\delta _{ij}
A_{i}(\vec{\boldsymbol{\sigma}})$ are different for the different
subspaces, with $A_{i}(\vec{\boldsymbol{\sigma}}_0) \leq
A_i(\vec{\boldsymbol{\sigma}})$ \cite{Note3}. Notice that if one
assigns $\sigma_{\alpha} = 0$ to the presence of a hole at site
$\alpha$, then $A_i(\vec{\boldsymbol{\sigma}}) = J_{z}
\sum_{\alpha=1}^{L-1} \sigma _{\alpha} \sigma _{\alpha +1}$. Therefore,
the Hamiltonian matrices H$_{i,j}^{{\cal M}}$ (of dimension ${\cal D}
\times {\cal D}$) in each subspace ${\cal M}$, consists of identical
off-diagonal matrix elements (H$_{i,j}^{{\cal M}}$ = H$_{i,j}^{{\cal
M^{\prime}}},\;i\neq j$) and different diagonal ones. These hermitian
matrices can be ordered according to the increasing value of the energy
$E_{M}$ of their parent states (for fixed $L$ and $M$). For any
$E_M(\vec{\boldsymbol{\sigma}}) < E_M(\vec{\boldsymbol{\sigma}}')$,
${\rm H}^{\cal M'} =  {\rm H}^{\cal M} + B$, where $B$ is a positive
semidefinite matrix. Then, the monotonicity theorem \cite{matrix} tells
us that 
\begin{equation}
E_{k}(\vec{\boldsymbol{\sigma}}) \leq
E_{k}(\vec{\boldsymbol{\sigma}}')\;\;\;\forall \ k=1,\cdots ,{\cal D} \ ,
\end{equation}
where $E_{k}(\vec{\boldsymbol{\sigma}})$'s are the eigenvalues of ${\rm
H}^{\cal M}$ arranged in increasing order
($E_{k}(\vec{\boldsymbol{\sigma}}) \leq
E_{k+1}(\vec{\boldsymbol{\sigma}})$). Therefore, we conclude that the
lowest eigenvalue of $\hat{H}$ must be in ${\cal M}_{0}$, is
$E_{1}(\vec{\boldsymbol{\sigma}}_0)$, and is two-fold degenerate.

{\it Spinless Fermion Mapping.}
The next step consists in showing, within the ground state
subspace ${\cal M}_{0}$, that the Hamiltonian $\hat{H}$ maps into an
attractive spinless fermion model. If one makes the following
identification 
\begin{equation}
|\underset{L-M}{\underbrace{\uparrow \downarrow \uparrow \downarrow
\cdots }} \ \underset{M}{\underbrace{\circ _{\;}\!\!\circ \circ \circ
\cdots }} \ \rangle \rightarrow | \underset{L-M}{\underbrace{\bullet_{\;}
\!\!\bullet \bullet \bullet \cdots }} \ \underset{M}{\underbrace{\circ_{\;}
\!\!\circ \circ \circ \cdots }} \ \rangle \ ,
\end{equation}
i.e., any spin ($c_{\alpha S}^{\dagger}$ or $c_{\alpha -S}^{\dagger}$)
maps into a single spinless fermion ($b_{\alpha }^{\dagger}$) in ${\cal
M}_{0}$, it is straightforward to show that all matrix elements of
H$^{{\cal M}_0}$ are identical to the matrix elements of the interacting
quantum lattice gas 
\begin{equation}
H=-t\sum_{\alpha=1}^{L-1} (b_{\alpha }^{\dagger }b_{\alpha +1}^{\;}+{\rm H.c.})
- J_{z}S^{2} \sum_{\alpha=1}^{L-1} n_{\alpha }n_{\alpha +1} 
\label{newH}
\end{equation}
in the corresponding new basis. In Eq. \ref{newH}, $n_{\alpha} =
b_{\alpha }^{\dagger} b_{\alpha }^{\;}$.

{\it Quantum Phase Diagram.}
The attractive spinless fermion model of Eq. \ref{newH} certainly has a
superconducting phase (i.e., correlation exponent $K_{\rho} > 1$)
\cite{our}. For arbitrary values of $J_{z}S^{2}$, $t$, and hole density
$\nu$, the spinless model is equivalent (via a Jordan-Wigner
transformation) to a Heisenberg-Ising spin-$\frac{1}{2}$ chain (also
known as XXZ model) with Hamiltonian \cite{Note4p}
\begin{equation} 
H_{\rm xxz} = \sum_{\alpha=1}^L J_{\perp} (s_{\alpha}^x
s_{\alpha+1}^x + s_{\alpha}^y s_{\alpha+1}^y) + J_{\|} 
(s_{\alpha}^z s_{\alpha+1}^z + s_{\alpha}^z) 
\label{xxz}
\end{equation} 
and $J_{\perp}=2t$, $J_{\|}=-J_{z}S^{2}$. In this new representation
the spin up(down) density is $\nu$($1-\nu$). $J_{\perp}=0$ represents
the classical Ising limit while $J_{\|}=0$ is the extreme quantum limit
(XY model). In general, the exchange anisotropy parameter $\Delta =
J_{\|}/J_{\perp} < 0$ determines the physical nature of the
correlations while $J_{\perp}$ defines the energy scale.

There is a vast literature on this model that was exactly solved by the
Bethe ansatz \cite{yan}. For $| \Delta | < 1$, solutions of this model
belong to the universality class called ``Luttinger liquids''
\cite{haldane}, with correlation functions characterized by power laws
with non-universal exponents continuously depending on $\Delta$.  The
correlation exponent $K_{\rho}$ is determined from 
\begin{equation}
K_{\rho} = \frac{\pi}{2} \nu^2 \kappa \, v_\rho = \pi \sqrt{\frac{\nu^2
\kappa D_{c}}{2}} \ ,
\end{equation}
where the Drude weight (or charge stiffness) $D_c$ is related to the
velocity of charge excitations $v_\rho$ by $D_c = v_\rho K_{\rho}/\pi$
with $D_{c} = \frac{1}{2} \partial^{2}e(\Phi)/\partial (\Phi/L)^{2}$ as
usual \cite{our}, and $\kappa$ is the isothermal compressibility. On
the other hand, $\kappa$ can be calculated from the variation of the
ground state energy per site $e(\nu)$ as $(\nu^2 \kappa)^{-1}=\partial^2
e/\partial \nu^2$.

At $\nu=1/2$, several quantities and properties are known in closed
analytic form. There is a Mott transition at $\Delta=1$ (Umklapp
scattering becomes relevant at $\Delta=1$, while it is irrelevant for
$| \Delta | < 1$). Moreover, the exact expressions for $K_{\rho}$ and
$v_\rho$ can be determined from the Bethe ansatz \cite{johnson}
\begin{equation}
K_{\rho}= \frac{\pi}{4 (\pi - \mu)} \; , \;\; v_\rho = \frac{\pi t \,
\sin \mu}{\mu} \ ,
\end{equation}
which implies $D_c=\pi t\sin \mu/[4\mu (\pi -\mu)]$ with $\cos \mu = 
\Delta$. The energy per site ($|\Delta| < 1$) is given by
\begin{equation}
e(1/2) = \frac{\Delta t}{2} - 2 t \sin{\mu} \int_0^\infty
\!\!\!\! dx \frac{\sinh{(\pi-\mu) x}}{\cosh{\mu x} \, \sinh{\pi x}} \ .
\end{equation}
Thus, one obtains $ \frac{1}{2} \leq K_{\rho} \leq 1$ in the
region $0 \geq \Delta \geq -1/\sqrt{2}$, and $K_{\rho}>1$
(superconducting correlations dominate at large distances) for $-1 <
\Delta < -1/\sqrt{2}$. At $\Delta=-1$, there is a transition to a phase
segregated state ($\kappa = 2 \mu /[\pi t \sin \mu \, (\pi - \mu)]$
diverges).

For $\Delta = 0$, the system reduces to a free spinless fermion system 
with energy per site $e=- \frac{2 t}{\pi} \sin(\pi \nu)$, stiffness
$D_c = -e/2$, and $\kappa^{-1} = - \pi^2 \nu^2 e$. This trivially
corresponds to $K_{\rho}=\frac{1}{2}$. Also the cases $\nu \rightarrow
0$ and $\nu \rightarrow 1$ map to free fermions independently of the
value of $\Delta$, therefore, $K_{\rho}=\frac{1}{2}$. The value
$\Delta = -1$ is also special: After the unitary transformation
$s_{\alpha}^{x,y} \rightarrow (-1)^\alpha s_{\alpha}^{x,y}$, the
Hamiltonian $H_{\rm xxz}$ maps into the FM Heisenberg model
in a magnetic field $J_{\|}$ (full $SU(2)$ symmetry is recovered). Here
$e=-\frac{t}{2} + \left ( \frac{1}{2} - \nu \right)$, implying the
opening of a gap with a diverging $\kappa$, i.e., $\Delta = -1$
determines the line of phase segregation for all densities $\nu$.
 
\vspace*{-2.5cm}
\begin{figure}[htb]
\epsfverbosetrue
\epsfxsize=12cm
$$
\centerline{\epsfbox{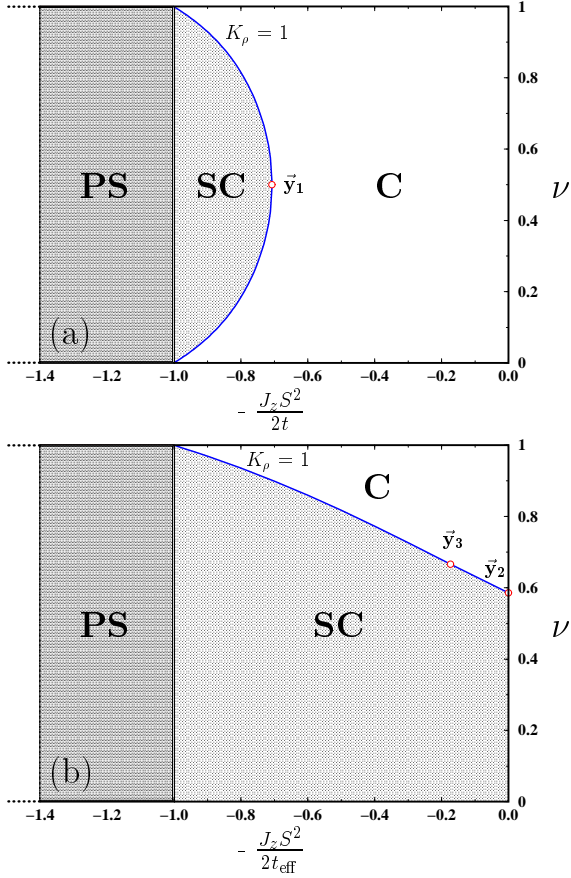}}
$$
\vspace*{-2.cm}
\caption{Quantum phase diagrams of the (a) $t$-$J_z$ chain, and (b)
modified $t$-$J_z$ model which includes the effects of the AF
background where the chain is embedded. There are three different
quantum phases: phase segregated (PS), superconducting (SC), and
metallic (C) phases. The last two belong to the ``Luttinger liquid''
universality class.  Points with $K_\rho=1$  known in closed analytic
form are: ${\bf \vec{y}_1} = (-\frac{1}{\sqrt{2}},\frac{1}{2})$, ${\bf
\vec{y}_2} = (0,2\!- \! \!\sqrt{2})$, and ${\bf \vec{y}_3} =
(\cos{\frac{5}{9}\pi},\frac{2}{3})$.}
\label{fig1}
\end{figure}

Away from $\nu=1/2$ and the special limiting cases discussed above, the
quantities $K_{\rho}$ and $v_\rho$ are obtained from the numerical
solution of the Bethe ansatz integral equations \cite{fowler,aligia}. To
calculate $v_\rho$ one needs to determine hole excitations with a
well-defined momentum $q$ and energy $\Delta e$ with respect to the
ground state $e$. We find the velocity of this elementary excitation
from $v_\rho= \lim_{q \rightarrow 0} d \Delta e/d q$, and together with
the numerical second order derivative of $e(\nu)$, we determine the
correlation exponent $K_\rho$. For $|\Delta| < 1$ the excitations are
gapless. The resulting quantum phase diagram is shown in Fig.
\ref{fig1}(a). Notice that the largest superconducting region
corresponds to $\nu=1/2$.

It is interesting to determine the influence of the antiphase domain
wall (ADW), associated with each charge, on the spin-spin correlations
of the metallic and the superconducting phases. It is well-known
\cite{structure} that  the charge structure factor of the spinless
model has a peak at $k=2 k_F = 2 \pi \nu$. Since each charge of the
effective model carries an ADW, the spin structure factor, ${\bf
S}(k)$, will peak at $k=\pi \pm 2 \pi \nu$ in the metallic phase. In
the superconducting phase a broad peak at $k=\pi$ is obtained for ${\bf
S}(k)$ since the pairs do not carry an ADW.

\begin{figure}[htb]
\epsfverbosetrue
\epsfxsize=8cm
$$
\centerline{\epsfbox{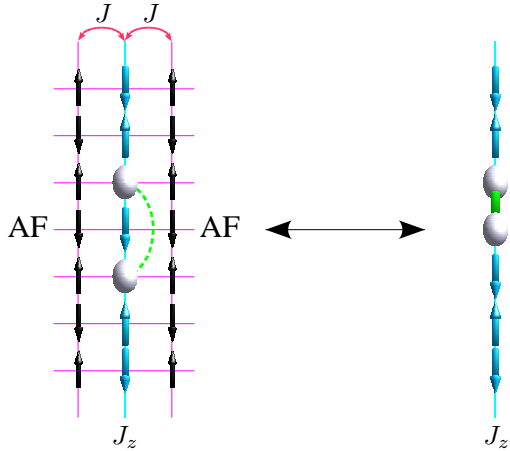}
\put(-31,0){\normalsize{$J_z$}}
\put(-171.5,0){\normalsize{$J_z$}}
\put(-178,160){\normalsize{$J$}}
\put(-160,160){\normalsize{$J$}}
}
$$
\caption{Schematic representation of the effect of an AF background on
a stripe segment. In this picture we assume that the concentration of
holes in the stripe is such that the there is no $\pi$ shift between AF
domains. If there were anti-phase domains, then, there is no confining
string potential even though two holes still like to be in adjacent sites. }
\label{fig2}
\end{figure}

{\it Effect of AF Boundaries on the Chain.}
It is also interesting to study the effect of the AF background in 
which our stripe segments are embedded. This background provides a
strong BC that results in an additional attractive (confining)
potential for the holes on the stripe. In this way, an enhanced
superconducting region is expected. In fact, the influence of the AF
background on the stripe is equivalent to the effect a staggered
magnetic field (STM) $B_s$ (see Fig. \ref{fig2}). Since each hole
carries an ADW, the staggered field gives rise to a confining linear
potential between the $\alpha$ and the $\alpha+1$ holes for even
$\alpha$. Therefore, the effect of a STM is to bind pairs of holes
tightly by a string potential. These pairs interact as hard core
bosons. In the very dilute limit one expects these bosons to condense
at $T=0$ for any value of $B_s/t$ and $J_z/t$ lower than some critical
value which gives rise to phase segregation.

In the limit $B_s/t \gg 1$ the model can be solved analytically for any
concentration of hole pairs. In this limit the problem reduces to
nearest-neighbors (NN) pairs of holes moving into an AF background (see
Fig. 2). Each pair can hop to its NN with an effective
hopping $t_{\rm eff}=2t^2/(B_s S + J_z S^2)$. In addition, there is an
attractive $J_z S^2$ interaction between NN pairs which comes from the
second term  of Eq. \ref{newH}. If we map each pair into a spinless
particle and each spin into an empty site \cite{cristian}
\begin{equation}
|\underset{(L,M)}{\underbrace{\circ \circ \uparrow \downarrow \uparrow
\downarrow \circ \circ \uparrow \downarrow \cdots }} \rangle \rightarrow |
\underset{(\tilde{L},\tilde{M})}{\underbrace{\bullet_{\;} \!\! \circ
\circ \circ \circ \bullet \circ \circ \cdots}} \ \rangle \ ,
\end{equation}
the problem reduces to the spinless Hamiltonian Eq. \ref{newH}. But as
each pair is replaced by an effective particle, the number of particles
and the length of the chain for the effective spinless problem are
\begin{equation}
\tilde{L}=L-\frac{M}{2}\;\;\;\;\;\;\;\;\;\tilde{M}=\frac{M}{2} \ ,
\label{newLN}
\end{equation}
where $\tilde{\nu} = \tilde{M}/\tilde{L} = \nu/(2-\nu)$. Here, as in
the FM case, we can use closed or open BC. The sign that arises after
a cyclic permutation of fermions is absorbed in the BC. For an
odd(even) number of fermions these BC are periodic(anti-periodic). This
corresponds to $M=4 n+2$ ($M=4 n$) of the original problem. The
relations between energies and charge velocities are \cite{Note5}
\begin{equation}
e(\nu) = \tilde{e}(\tilde{\nu}) \left (1- \frac{\nu}{2} \right )\; , \;\;
v_\rho(\nu) = \tilde{v}_\rho(\tilde{\nu}) \frac{2}{2 - \nu} \ ,
\end{equation}
where $\tilde{e}(\tilde{\nu})$ and $\tilde{v}_\rho(\tilde{\nu})$ are
the energy per site and charge velocity of the corresponding spinless
model of concentration $\tilde{v}$. Therefore, simple algebraic
manipulations lead to 
\begin{equation}
K_\rho(\nu) = \tilde{K}_\rho(\tilde{\nu}) \ (2 - \nu)^2 \ ,
\end{equation}
and the phase diagram is depicted in Fig. \ref{fig1}(b). 

\vspace*{-0.6cm}

For completeness, we would like to mention that the mapping of the low
energy spectra of the $t$-$J_z$ model into the spinless Hamiltonian,
Eq. \ref{newH}, is also valid for the FM case, i.e., $J_z < 0$. In this
case, the magnetic background, which is replaced by empty sites in the
spinless model, is FM. Notice, however, that the effective spinless
model is also attractive ($\Delta < 0$). This implies that the dynamics
of  the charge degrees of freedom in an AF background is the same as in
the FM one. But in the latter case, the charges do not carry an ADW.
Moreover, the mapping does not depend upon the statistics of the
quantum particles. In other words, we could also apply these concepts
to constrained quantum particles with integer spin $S$, i.e., hard-core
bosons. In the large spin $S$ limit, the quantum phase diagram of the
$t$-$J_z$ model approaches the one of isotropic $t$-$J$ Hamiltonian.
Notice, however, the qualitative similarity between the phase diagram
in Fig.~\ref{fig1}(a) and the one for the isotropic spin-$\frac{1}{2}$
$t$-$J$ model obtained numerically \cite{ogata}. Finally, the solution
can be trivially extended to the $t$-$J_z$-$V$ model, where $V$
represents a NN density-density interaction. The effect of $V$ is
simply to renormalize the spinless fermion interaction in Eq.
\ref{newH}. Furthermore, it is simple to prove that there is a family
of bilinear-biquadratic spin-1 chain Hamiltonians that can be mapped
onto a $t$-$J_z$-$V$ model and, therefore, its low energy physics is
exactly solvable \cite{cristian1}. 

In summary, we presented the exact quantum phase diagram of the
$t$-$J_z$ chain model for arbitrary spin $S$, particle statistics, and
sign of the magnetic interaction $J_z$. We also exactly determined the
phase diagram of a modified $t$-$J_z$ chain that includes the effects
of a strong antiferromagnetic background. A metallic, superconducting
and segregated phases characterize these two phase diagrams. 

We thank A.A. Aligia, S. Sondhi, and J.E. Gubernatis for useful
discussions, and A.A. Aligia for helping us with the Bethe ansatz
equations. Work at Los Alamos is sponsored by the US DOE under contract
W-7405-ENG-36.

\end{multicols}

\end{document}